\documentclass[journal,10pt]{IEEEtran}

\usepackage[dvips]{graphicx}
\usepackage{amsmath,amsfonts,amssymb}

\usepackage{subeqnarray}
\usepackage{cases}

\usepackage{epsfig}
\usepackage{fixfoot}
\usepackage{stfloats, balance}
\usepackage{algorithm}
\usepackage{algorithmic}
\usepackage{xcolor}

\IEEEoverridecommandlockouts

\title{Double-Target Collaborative Spectrum Sharing for 6G Hybrid Satellite-Terrestrial Networks with {User-Centric Channel Pools}}
\author{{Yanmin Wang, Wei Feng, \emph{Senior Member, IEEE}, Ming Xiao, \emph{Senior Member, IEEE}, \\Cheng-Xiang Wang, \emph{Fellow, IEEE}}
\thanks{
Y. Wang is with the School of Information Engineering, Minzu University of China, Beijing 100081, China~(email: wangyanmin@muc.edu.cn).

W. Feng is with the Department of Electronic
Engineering, Tsinghua University, Beijing 100084, China~(e-mail:
fengwei@tsinghua.edu.cn).

M. Xiao is with the Communication Theory Laboratory, KTH Royal
Institute of Technology, 10044 Stockholm, Sweden (e-mail: mingx@kth.se).

C.-X. Wang (corresponding author) is with the National Mobile Communications Research Laboratory,
School of Information Science and Engineering, Southeast University,
Nanjing 210096, China. He is also with the Purple Mountain Laboratories,
Nanjing 211111, China (e-mail: chxwang@seu.edu.cn).
}
}

\begin{document}
\maketitle

\begin{abstract}
Satellite and terrestrial cellular networks can be integrated together for extended broadband coverage 
in e.g., maritime communication scenarios, in the upcoming sixth-generation (6G) era.
{To counter spectrum scarcity, collaborative spectrum sharing 
	is considered for a hybrid satellite-terrestrial network (HSTN) in this paper.
	With only slowly-varying large-scale channel state information (CSI),
	joint power and channel allocation is implemented for terrestrial mobile terminals (MTs) which share the same frequency band 
	with the satellite MTs opportunistically.
	Specially, strict quality service assurance is adopted for terrestrial MTs 
	under the constraint of leakage interference to satellite MTs.
	With the target of maximizing both the number of served terrestrial MTs 
	and the average sum transmission rate,
	a double-target spectrum sharing problem is formulated.
	To solve the complicated mixed integer programming (MIP) problem efficiently,
	user-centric channel pools are introduced.
	Simulations demonstrate that the proposed spectrum sharing scheme could 
	achieve a significant performance gain for the HSTN.}
\end{abstract}




\section{introduction}
\label{s1}
Currently, the terrestrial fifth-generation (5G) network is able to provide a high communication rate, but its
coverage performance crucially depends on densely-deployed base stations (BSs). In rural or maritime areas without sufficient available BS cites, the
broadband coverage region of terrestrial networks is usually quite limited~\cite{r1, 10466703}. Thereby, satellite communications can be integrated
for extended broadband coverage in the upcoming sixth-generation (6G) era, leading to a hybrid satellite-terrestrial network (HSTN)~\cite{r5, 10477428}.

In an HSTN, spectrum may be shared between satellite and terrestrial systems, to alleviate spectrum scarcity. 
{When full channel state information (CSI) is available, a significant performance gain can be achieved 
	with effective co-channel interference (CCI) mitigation~\cite{rtwc20172,rJSTSP2019,rtwc2022}.
	In~\cite{rtwc20172}, the outage probability was derived for a hybrid satellite-terrestrial spectrum sharing system.
	A joint beamforming and power allocation scheme was proposed in~\cite{rJSTSP2019} to maximize
	the sum rate of satellite-terrestrial integrated networks.
	In~\cite{rtwc2022}, with the presence of a primary satellite-receiver link,
	multiple UAVs with aerial stations and a terrestrial BS were deployed to support smart vehicles.
	However, due to large propagation delay of the satellite links, it is rather challenging to obtain fast-varying full CSI
	for satellite-terrestrial spectrum sharing in an HSTN.}

{Compared to full CSI, statistical or large-scale CSI usually varies much more slowly
	and can be acquired with the aid of offline-obtained data in many circumstances~\cite{r21,10679984,10038856}. 	
	To circumvent the challenge of full CSI acquisition, 
	researchers have turned to spectrum sharing based on statistical or large-scale CSI~\cite{8331998, 1111111, 10678835}.
	With only information of path loss and shadowing obtained from a pre-constructed radio map~\cite{10371362},
	power allocation and user scheduling schemes were proposed for a hybrid network 
	with satellite-terrestrial spectrum sharing in~\cite{8331998}.
	With statistical or instantaneous interference constraints imposed by primary terrestrial links,
	energy efficient power allocation was investigated for cognitive satellite-terrestrial networks in~\cite{1111111}.
	A fine-over-coarse spectrum sharing scheme with shaped virtual cells is proposed
	for hybrid satellite-UAV-terrestrial maritime networks based on joint subcarrier and time slice allocation in~\cite{10678835}.
	\cite{8331998, 1111111, 10678835} indicate that 
	the performance gain is still elusive when only statistical or large-scale CSI is known.}

{In this paper, we focus on large-scale-CSI-based collaborative spectrum sharing for an HSTN,
	where terrestrial mobile terminals (MTs) share the same frequency band with satellite MTs opportunistically.
	Specifically, joint power and channel allocation is implemented for terrestrial MTs
	under the constraint of leakage interference to satellite MTs.
	Different from previous work~\cite{8331998, 1111111, 10678835},
	strict quality service assurance is adopted for the terrestrial MTs.
	That is only terrestrial MTs whose service quality can be satisfied could be served via spectrum sharing.
	With the target of maximizing both the number of served terrestrial MTs and the average sum transmission rate,
	a double-target spectrum sharing problem is formulated.
	To solve the complicated mixed integer programming (MIP) problem efficiently,
	the user-centric channel pools are introduced and 
	a joint power and channel allocation scheme is proposed.}

\section{System Model and Problem Formulation}
\begin{figure}[t]
	\begin{center}
		\includegraphics[width=6.6 cm]{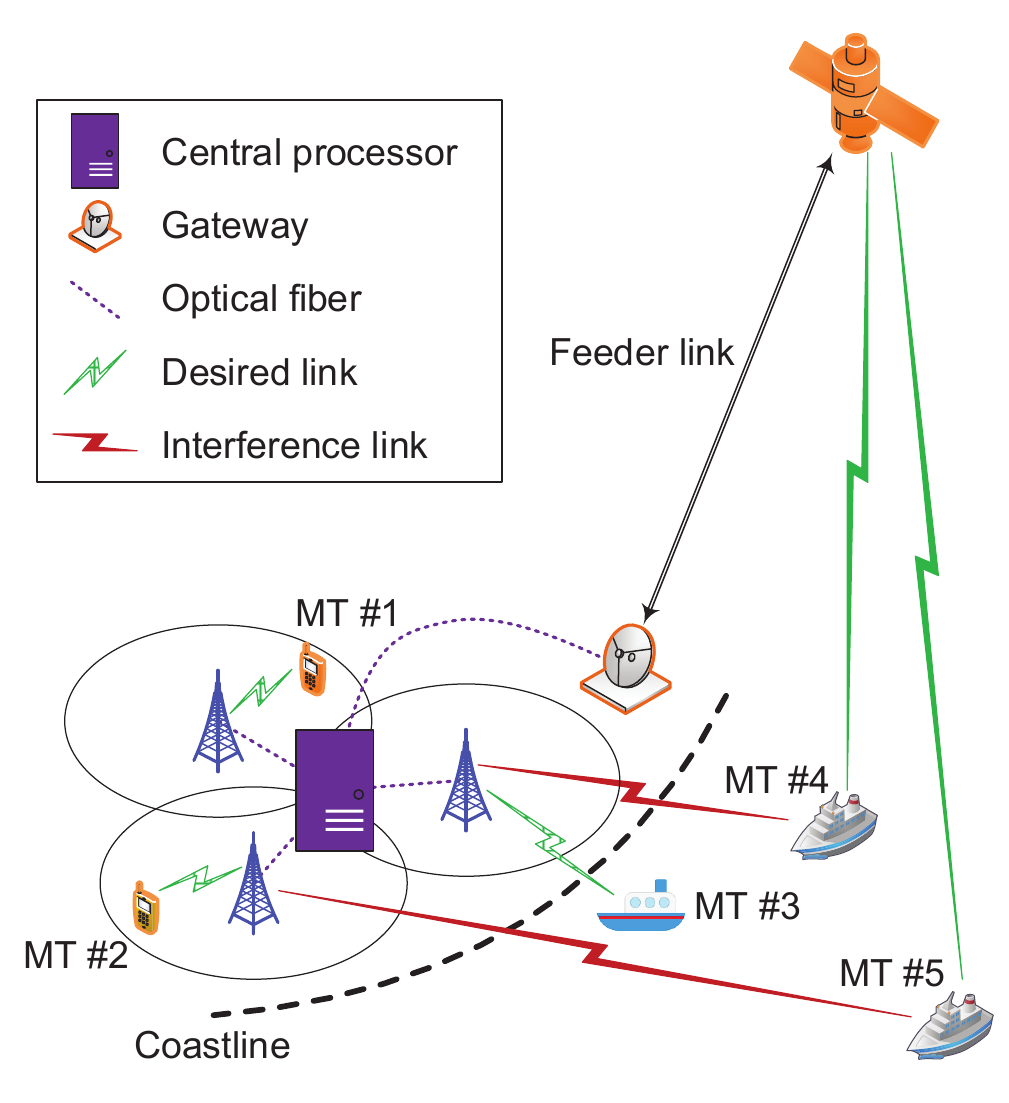}
		\caption{Illustration of an HSTN: a coastwise case.}
		\label{fig1}
	\end{center}
\end{figure}

As illustrated in Fig.~\ref{fig1}, we consider a {coastwise} HSTN consisting of a satellite, $K_s$ satellite MTs,
$N$ terrestrial BSs, and $K_t$ terrestrial MTs equipped with $M$ antenna elements each.
{The satellite MTs are assumed to be all ship-borne and served in the downlink by the satellite in $K_s$ orthogonal channels,
	each with a bandwidth of $B$.
	For higher spectrum efficiency, the terrestrial system shares the same frequency band with the satellite part opportunistically.
	All the BSs are connected to a central processor and serve the terrestrial MTs cooperatively in the downlink.	
	The terrestrial MTs can be either ship-borne or land-based.
	To enable coordination between the satellite and terrestrial systems,
	the satellite gateway is also connected to the terrestrial central processor, e.g., by optical fibers.}

{Without loss of generality, it is assumed that satellite MT $j$ is served in channel $j$, $j = 1,2,...,K_s$,
	and $K_s \geq K_t$. 
	To avoid harmful CCI, the maximum average interference suffered by $K_s$ satellite MTs from the terrestrial system 
	is restricted by $\mathfrak{I}^t_{j}$, $j = 1,2,...,K_s$, respectively.
	Besides, each channel can be shared at most by one terrestrial MT, 
	and each terrestrial MT $i$, $i=1,...,K_t$, is served at most in one channel.
	Meanwhile, to ensure service quality of the terrestrial system,
	it is required that a minimum average transmission rate $\underline{R}_{i}$ must be achieved 
	for terrestrial MT $i$ once it is served in one of the channels.}
To promote spectrum sharing performance,
we optimize both channel and power allocation for the terrestrial MTs. 
{As the target, 
	the number of terrestrial MTs that are served with service quality assurance is maximized firstly, and then 
	the average sum transmission rate of the terrestrial system is maximized.}

{Let $z_{ij}$, $i=1,...,K_t$, $j = 1,2,...,K_s$, 
	denote channel allocation indicators for $K_t$ terrestrial MTs in $K_s$ channels.}
If terrestrial MT $i$ is scheduled in channel $j$, $z_{ij}=1$. Otherwise, $z_{ij}=0$. 
{Further,} let 
$\mathbf{P}_{ij}= \text{diag} \{p_{ij1},p_{ij2},...,p_{ijN}\}$
denote the transmit power matrix for terrestrial MT $i$ in channel $j$ {from $N$ BSs}. 
{With a total transmit power constraint $P_i$ for each terrestrial MT $i$, }
the {double-target spectrum sharing} problem can be formulated as
\begin{subequations} \label{eq1}
	\begin{align}
		\max_{{ \{  z_{i,j}, \mathbf{P}_{ij} \} } } & { -  \mathcal{M}( K_t - \sum_{i=1}^{K_t} \sum_{j=1}^{K_s}  z_{i,j} )  +} \sum_{i=1}^{{K_t}} \sum_{j=1}^{{K_s}} { z_{i,j} R_{ij} B}  \\
		\text{s.t.}~ & \sum_{j=1}^{{K_s}} z_{ij} \sum_{n=1}^N p_{ijn} \leq P_{i},\, i = 1,2,...,{K_t,} \\
		& {\sum_{j=1}^{K_s}  z_{i,j} R_{ij} B \geq \underline{R}_{i} \sum_{j=1}^{K_s}  z_{i,j} , \, i = 1,2,...,K_t,} \\
		&  \sum_{i=1}^{{K_t}} z_{ij} \sum_{n=1}^N p_{ijn} (l_{jn}^{s})^2 \leq \mathfrak{I}_j^t, \, j = 1,2,...,{K_s}, \\
		& \sum_{i=1}^{{K_t}} z_{ij} {\leq} 1, \, \sum_{j=1}^{{K_s}} z_{ij} {\leq} 1,  \, {\forall i,j,} \\
		& p_{ijn} \geq 0, z_{ij} \in \{0, 1\}, \, \forall i,j,n.
	\end{align}
\end{subequations}
where
\begin{eqnarray} \label{eq1_1}
	R_{ij} = \mathbb{E} [ \log_2\det (\mathbf{I}_M+ \frac{ \mathbf{H}_{ij}\mathbf{P}_{ij}\mathbf{H}_{ij}^H}{ \mathfrak{I}_{ij}^s +\sigma^2}) ],
\end{eqnarray}
is the average transmission rate per Hz for terrestrial MT $i$ in channel $j$.
{The constraints in~(\ref{eq1}c) 
	indicate that each terrestrial MT $i$ is either served with service quality assurance
	or not served in $K_s$ shared channels at all.}
{$\mathcal{M}$ is a large-enough positive number for punishment on cases 
	when there are terrestrial MTs not served via spectrum sharing, 
	i.e., $\sum_{i=1}^{K_t} \sum_{j=1}^{K_s}  z_{i,j} < K_t$.
	It can be inferred that
	\begin{eqnarray} \label{eq1_2}
		\mathcal{M} \geq B \sum_{i=1}^{K_t} \max_{j \in \{1,...,K_s\}} [ \max_{\mathbf{P}_{ij}} R_{ij} ],
	\end{eqnarray}
	could guarantee priority of the number of terrestrial MTs served with service quality assurance over 
	the average sum transmission rate of the terrestrial system in maximization.}
$\mathbb{E} {[\cdot]}$ denotes the expectation operator with respect to the unknown small-scale channel fading,
$\mathfrak{I}_{ij}^s$ is interference from the satellite to terrestrial MT $i$ in channel $j$,
and $\sigma^2$ denotes power of the white Gaussian noise.
$\mathbf{H}_{ij} {\in \mathbb{C}^{M \times N}}$ represents the channel {from $N$ BSs} to terrestrial MT $i$ in channel $j$,
and $\mathbf{h}_{j} {\in \mathbb{C}^{1 \times N}}$ represents the channel from $N$ BSs to satellite MT $j$, which can be respectively expressed as~\cite{8331998, 10678835, r15}
\begin{subequations}\label{eq2}
	\begin{align}
		&\mathbf{H}_{ij} = \mathbf{S}_{ij}\left[ \begin{array}{ccc}
			l_{ij1}^{t} &  &  \\
			& \ddots &  \\
			&  & l_{ijN}^{t}
		\end{array} \right],\\
		&\mathbf{h}_{j} =
		\mathbf{s}_{j}\left[ \begin{array}{ccc}
			l_{j1}^{s} &  &  \\
			& \ddots &  \\
			&  & l_{jN}^{s}
		\end{array} \right].
	\end{align}
\end{subequations}
$l_{ijn}^{t},\,n = 1,2,...,N,$ and $l_{jn}^{s},\,n = 1,2,...,N,$ denote the large-scale fading from terrestrial BS $n$ to terrestrial MT $i$ and satellite MT $j$ in channel $j$, respectively.
$\mathbf{S}_{ij} {\in \mathbb{C}^{M \times N}}$ and $\mathbf{s}_{j} {\in \mathbb{C}^{1 \times N}}$ 
denote the small-scale Rayleigh fading.
While $\mathbf{S}_{ij}$ and $\mathbf{s}_{j}$ are usually fast-varying and difficult to obtain in practice,
$l_{ijn}^{t}$ and $l_{jn}^{s}$ usually vary much more slowly and can be obtained from historical data 
or via a very small amount of system overhead~\cite{8331998, 10678835, r15}.

{The spectrum sharing problem in~(\ref{eq1}) is a complicated
	mixed integer programming (MIP) problem that is challenging to solve.
	The expectation operator $\mathbb{E} [\cdot]$ in the objective function,
	as shown in~(\ref{eq1}a) and~(\ref{eq1_1}), further complicates the situation.}

{\section{Spectrum Sharing with User-Centric Channel Pools}}
{It can be observed that the optimal channel allocation indicators $z_{ij}$, $\forall i,j$, for~(\ref{eq1}) 
	can be determined once all the maximum achievable $R_{ij}$, $\forall i,j$, are obtained.}
In what follows, the problem in~(\ref{eq1}) is solved via a divide-and-conquer approach, i.e.,
implementing power and channel allocation successively.
{Specially, to execute channel allocation efficiently after power allocation,
	user-centric channel pools are formed for the terrestrial MTs to indicate available channels for each of them 
	with respect to constraints on both CCI and service quality assurance.}

{Firstly, the maximum achievable $R_{ij}$ for each terrestrial MT in each channel
	is obtained via power allocation.
	Based on~(\ref{eq1}),} the power allocation subproblem for terrestrial MT $i$ in channel $j$ can be written as
\begin{subequations} \label{eq4}
	\begin{align}
		\max_{{\mathbf{P}_{ij}}} \,\,  & R_{ij} B \\
		\text{s.t.}~& \sum_{n=1}^N p_{ijn} \leq P_{i}, \\
		&   \sum_{n=1}^N p_{ijn} (l_{jn}^{s})^2 \leq \mathfrak{I}_j^t, \\
		& p_{ijn} \geq 0, \, \forall i,j,n.
	\end{align}
\end{subequations}
{With given $\mathfrak{I}_{ij}^s$, (\ref{eq4}) is actually a convex problem~\cite{r15}.}
The key difficulty of solving the problem lies in the expectation operator $\mathbb{E}$ in (\ref{eq4}a), 
which actually requires complicated integral operation. 
{Fortunately, it has been proved that (\ref{eq4}) can be transformed into}
a max-min problem {with a concave-convex objective function}, which can be efficiently solved~\cite{r15, r16}.

Specifically, based on the random matrix theory, 
an accurate closed-form approximation can be obtained for $R_{ij}$ given in~(\ref{eq1_1}), as~\cite{r15, r14}
\begin{eqnarray} \label{eq5}
	&&\!\!\!\!\!\!\!\!\!\!\!\!\!\!\!\!\! \tilde{R}_{ij} = \sum_{n=1}^N \log_2(1+\frac{p_{ijn} (l_{ijn}^{t})^2 M}{(\mathfrak{I}_{ij}^s +\sigma^2) \chi_{ij} } ) + M \log_2(\chi_{ij}) \nonumber \\
	&&\!\! - M\log_2 e (1 - \chi_{ij}^{-1}),
\end{eqnarray}
where $\chi_{ij}$ satisfies
\begin{eqnarray}\label{eq6}
	\chi_{ij}=1+\sum_{n=1}^N\frac{p_{ijn} (l_{ijn}^{t})^2 \chi_{ij}} {(\mathfrak{I}_{ij}^s +\sigma^2) \chi_{ij}+ p_{ijn} (l_{ijn}^{t})^2
		M}.
\end{eqnarray}
Define 
\begin{eqnarray} \label{eq7}
	&&\!\!\!\!\!\!\!\!\!\!\!\!\!\!\!\!\! y_{ij}(\mathbf{P}_{ij}, x_{ij}) = \sum_{n=1}^N \log_2(1+\frac{p_{ijn} (l_{ijn}^{t})^2 M}{(\mathfrak{I}_{ij}^s +\sigma^2) e^{x_{ij}} } ) \nonumber \\
	&&\!\! + M\log_2 e [x_{ij} + e^{-x_{ij}} - 1],
\end{eqnarray}
and then we have~\cite{r15}
\begin{eqnarray}\label{eq8}
	\tilde{R}_{ij}= \underset{ x_{ij} \geq 0} \min \, y_{ij}(\mathbf{P}_{ij}, x_{ij}).
\end{eqnarray}	
By replacing $R_{ij}$ with $\tilde{R}_{ij}$ and removing the constant $B$ in~(\ref{eq4}), the problem can be transformed into 
a max-min problem as
\begin{subequations} \label{eq9}
	\begin{align}
		\max_{\mathbf{P}_{ij}} \min_{x_{ij}} \,\,  & y_{ij} (\mathbf{P}_{ij}, x_{ij}) \\
		\text{s.t.}~& (\ref{eq4}b), (\ref{eq4}c), (\ref{eq4}d), \nonumber \\
		& x_{ij} \geq 0.
	\end{align}
\end{subequations}
As $y_{ij}(\mathbf{P}_{ij}, x_{ij})$ is concave with respect to $\mathbf{P}_{ij}$ and convex with respect to $x_{ij}$,
(\ref{eq9}) can be solved based on standard algorithms~\cite{r15, r16}.

\begin{figure}[t]
	\begin{center}
		\includegraphics[width=8.3cm]{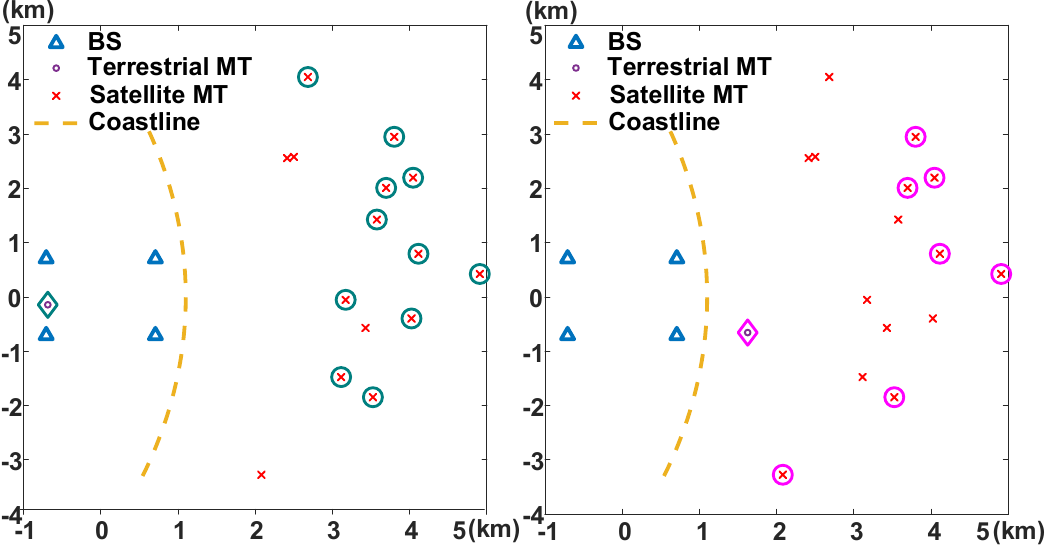}
		\caption{Illustration of the user-centric channel pools.}
		\label{fig2}
	\end{center}
\end{figure}

{Let $\mathbf{P}^*_{ij}$, $\forall i,j$, denote the optimal transmit power matrices for~(\ref{eq9}), 
	and $R^*_{ij}$, $\forall i,j$, are the corresponding maximum average transmission rates.
	With $R^*_{ij}$, $\forall i,j$, (\ref{eq1}) can be transformed into a channel allocation subproblem as
	\begin{subequations} \label{eq10}
		\begin{align}
			\max_{ \{  z_{i,j} \} } & -  \mathcal{M}( K_t - \sum_{i=1}^{K_t} \sum_{j=1}^{K_s}  z_{i,j} ) + \sum_{i=1}^{K_t} \sum_{j=1}^{K_s}  z_{i,j} R^*_{ij} B \\
			\text{s.t.}~ & \sum_{j=1}^{K_s}  z_{i,j} R^*_{ij} B \geq \underline{R}_{i} \sum_{j=1}^{K_s}  z_{i,j}, \, i = 1,2,...,K_t, \\
			& \sum_{i=1}^{K_t} z_{ij} \leq 1, \, \sum_{j=1}^{K_s} z_{ij} \leq 1, z_{ij} \in \{0, 1\}, \, \forall i,j.
		\end{align}
	\end{subequations}
	Because of the constraints in~(\ref{eq10}b) as well as the inequality in~(\ref{eq10}c), 
	there is not a straightforward low-complexity algorithm for the assignment problem in~(\ref{eq10}).}
{For efficient spectrum sharing, we introduce user-centric channel pools
	to indicate available channels for each terrestrial MT for channel allocation, as
	\begin{eqnarray} \label{eq11}
		\mathbb{G}_{i} = \{j \,|\, R^*_{ij} B \geq \underline{R}_{i}, j=1,...,K_s \}, i=1,...,K_t. 
	\end{eqnarray}
	It could be inferred from~(\ref{eq4}), (\ref{eq9}) and~(\ref{eq11}) that $\mathbb{G}_{i}$ represents the subset of channels 
	that could be shared by terrestrial MT $i$ with respect to both CCI constraints and service quality assurance.
	Fig.~\ref{fig2} shows the user-centric channel pools for two of the terrestrial MTs 
	under a random topology of the HSTN considered in Section IV.
	In each sub-figure, the terrestrial MT is indicated by a diamond 
	and satellite MTs served in the available channels for it are indicated by circles.}

With the aid of user-centric channel pools $\mathbb{G}_{i}$ given in~(\ref{eq11}),
	we try to transform~(\ref{eq10}) into a standard assignment problem that can be efficiently solved~\cite{r17}. 
	Specially, for $i, j = 1,...,K_s$, set 
	\begin{equation}\label{eq13}
		{R'}^*_{ij} = \left \{
		\begin{aligned}
			& R^*_{ij}, \,\, \,\,\,\,\,\, i \in\{1,...,K_t\} \, \text{and} \, j \in \mathbb{G}_{i},\\
			& -\mathcal{M}/B, \,\,  \text{otherwise},
		\end{aligned}
		\right.
	\end{equation}
	and formulate a standard assignment problem as  
\begin{subequations} \label{eq12}
	\begin{align}
		\max_{ \{  z'_{i,j} \} } & \sum_{i=1}^{K_s} \sum_{j=1}^{K_s}  z'_{i,j} {R'}^*_{ij} \\
		\text{s.t.}~ & \sum_{i=1}^{K_s} z'_{ij} = 1, \, \sum_{j=1}^{K_s} z'_{ij} = 1, z'_{ij} \in \{0, 1\}, \nonumber \\
		& \,\,\,\,\,\,\,\,\,\,\,\,\,\,\,\,\,\,\,\,\,\,\,\,\,\,\,\,\,\,\,\,\,\,\,\,\,\,\,\,\,\,\,\,\,\,\,\,\,\,\,\,\,\,\,\,\, i,j = 1,...,K_s.
	\end{align}
\end{subequations}
It can be solved by the Kuhn-Munkres algorithm with a complexity of $O(K_s^3)$~\cite{r17}.
{Suppose ${z'}^*_{i,j}$, $i, j = 1,...,K_s$, constitute an optimal solution for~(\ref{eq12}).
	With $\mathcal{M} \geq B \sum_{i=1}^{K_t} \max_{j \in \{1,...,K_s\}} R^*_{ij}$, as shown by~(\ref{eq1_2}),
	it could be proved that , 
	\begin{eqnarray} \label{eq15}
		z^*_{i,j} = {z'}^*_{i,j}, \, i = 1,...,K_t, j=1,...,K_s,
	\end{eqnarray}
	are optimal for~(\ref{eq10}).}

Based on the divide-and-conquer approach illustrated above, 
we propose a spectrum sharing scheme, as shown in Algorithm~\ref{tab1}.
For execution of the proposed scheme, $K_t K_s$ max-min subproblems as shown in~(\ref{eq9}) 
and one standard assignment problem as shown in~(\ref{eq12}) need to be solved. 
{As~(\ref{eq9}) can be solved with a polynomial complexity with respect to $N$ and 
	$N$ is usually very small in practical applications, 
	it could be inferred that the overall complexity of Algorithm~\ref{tab1} is $O(K_s^3)$.}
Simulations in Section IV demonstrate that
the proposed spectrum sharing scheme can achieve a significant performance gain.

\begin{algorithm}[t]
	\caption{Proposed spectrum sharing scheme.}
	\label{tab1}
	\begin{algorithmic}[1]
		\FOR{$i=1$ to $K_t$}
		\FOR{$j=1$ to $K_s$}
		\STATE Solve the power allocation subproblems in~(\ref{eq9}), and obtain $R^*_{ij}$, $i=1,...,K_t$, $j=1,...,K_s$.
		\ENDFOR
		\ENDFOR
		\STATE Determine the user-centric channel pools $\mathbb{G}_{i}$, $i=1,...,K_t$, as shown in~(\ref{eq11}).
		\STATE Solve the channel allocation subproblem in (\ref{eq12}) and derive ${z'}^*_{ij}$,  $\,i,\,j=1,...,K_s$.
		\STATE Output: $z^*_{ij} = {z'}^*_{ij},~i=1,2,...,K_t,~j=1,2,...,K_s$.
	\end{algorithmic}
\end{algorithm}

\begin{figure}[t]
	\begin{center}
		\includegraphics[width=7 cm]{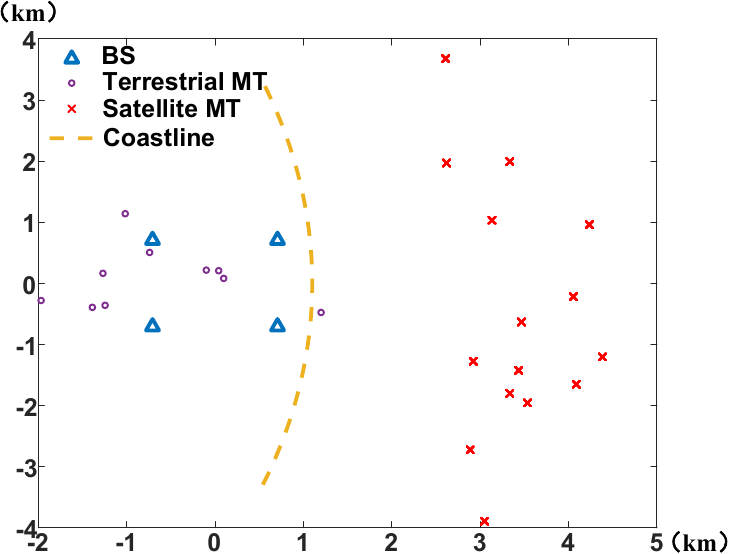}
		\caption{A randomly-generated system topology of the HSTN for simulations.}
		\label{fig3}
	\end{center}
\end{figure}

\section{Simulation results}
As shown in Fig.~\ref{fig3}, we consider {an HSTN with $N = 4$, $K_t = 10$, $M = 4$, and $K_s = 15$.
	The BSs are deployed by a coastline. The terrestrial MTs are randomly distributed in a circular area with a radius of $2$km around the BSs.
	As mentioned in Section II, the terrestrial MTs could be either ship-borne or land-based, depending on their specific location.
	The ship-borne satellite MTs are assumed to be randomly distributed in a $150^{\circ}$ sector of a larger circular area 
	with a radius of $5$km.
	The distance between each satellite MT and the center of the circular area is assumed to be larger than $3$km.}

{The bandwidth of each channel is set as $B=1$MHz, and the center frequency of the channels is set as $f_c=2$GHz.
	A general shadowed path loss model $l^2 [dB]=\text{FSPL}(f_c,d_0)+ a\log_{10}(d/d_0) + \gamma$ 
	is adopted for the large-scale fading $l_{ijn}^{t}$ and $l_{jn}^{s}$, $\forall i,j,n$.
	In the model, $d_0=1$m denotes the reference distance, $d$ is the transmission distance, 
	$\text{FSPL}(f_c,d_0)$ is the free space path loss at $d_0$,
	$a$ represents the path loss exponent, and $\gamma$ denotes the lognormal-distributed shadowing~\cite{7434656}.
	In our simulations, $a$ is set as $3$, and the standard deviation of the shadowing is $4$dB~\cite{7434656}.
	The total transmit power constraint for each terrestrial MT is set as $P_i \in [10,30]$dBm, $\forall i$,
	and the noise power is $\sigma^2 = -114$ dBm.
	The threshold for interference suffered by the satellite MTs is set as $\mathfrak{I}_j^t= \sigma^2-12.2\text{dB}-G_r$,
	$j=1,...,K_s$, where $G_r$ is the receive antenna gain and $12.2$dB is the protection criterion 
	set by the International Telecommunication Union (ITU) for satellite communication systems~\cite{ITURM1799}.
	When interference is received at the antenna sidelobe with $G_r = -10$dB~\cite{10678835}, 
	it could be obtained that $\mathfrak{I}_j^t = -116.2$dBm.
	The interference suffered by the terrestrial MTs from the satellite
	is assumed to be $\mathfrak{I}_{ij}^s = -119$dBm, $i=1,...,K_t$, $j=1,...,K_s$.
	Note that $-119$dBm is obtained from a typical power flux density at Starlink user terminals, which is $-146$dBW/$m^2$/$4$kHz,
	based on the free-space satellite-terrestrial propagation~\cite{Starlink2023}.
	The service quality for the terrestrial MTs is set as $\underline{R}_{i} = \alpha R^*_{ij} B $, $i=1,...,K_t$,
	with $\alpha = 1/10, 1/3, 1/2$.}

\begin{figure}[t]
	\begin{center}
		\includegraphics[width=7 cm]{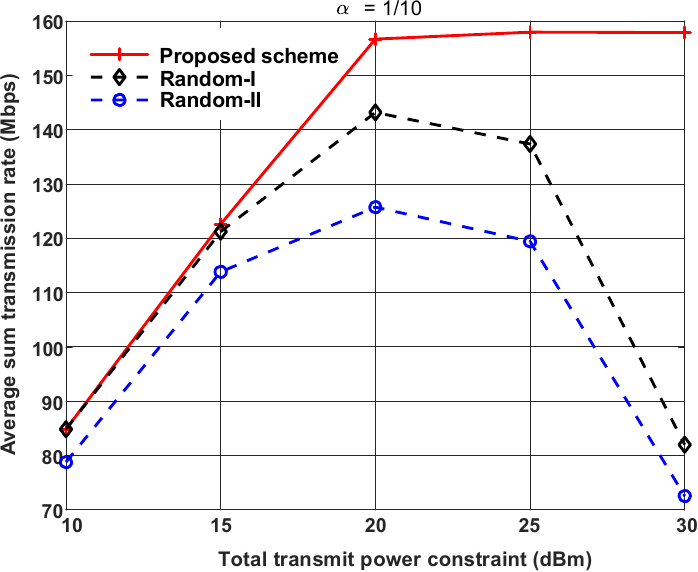}
		\caption{Average sum transmission rate for $\alpha = 1/10$.}
		\label{fig4}
	\end{center}
\end{figure}

\begin{figure}[t]
	\begin{center}
		\includegraphics[width=8.3 cm]{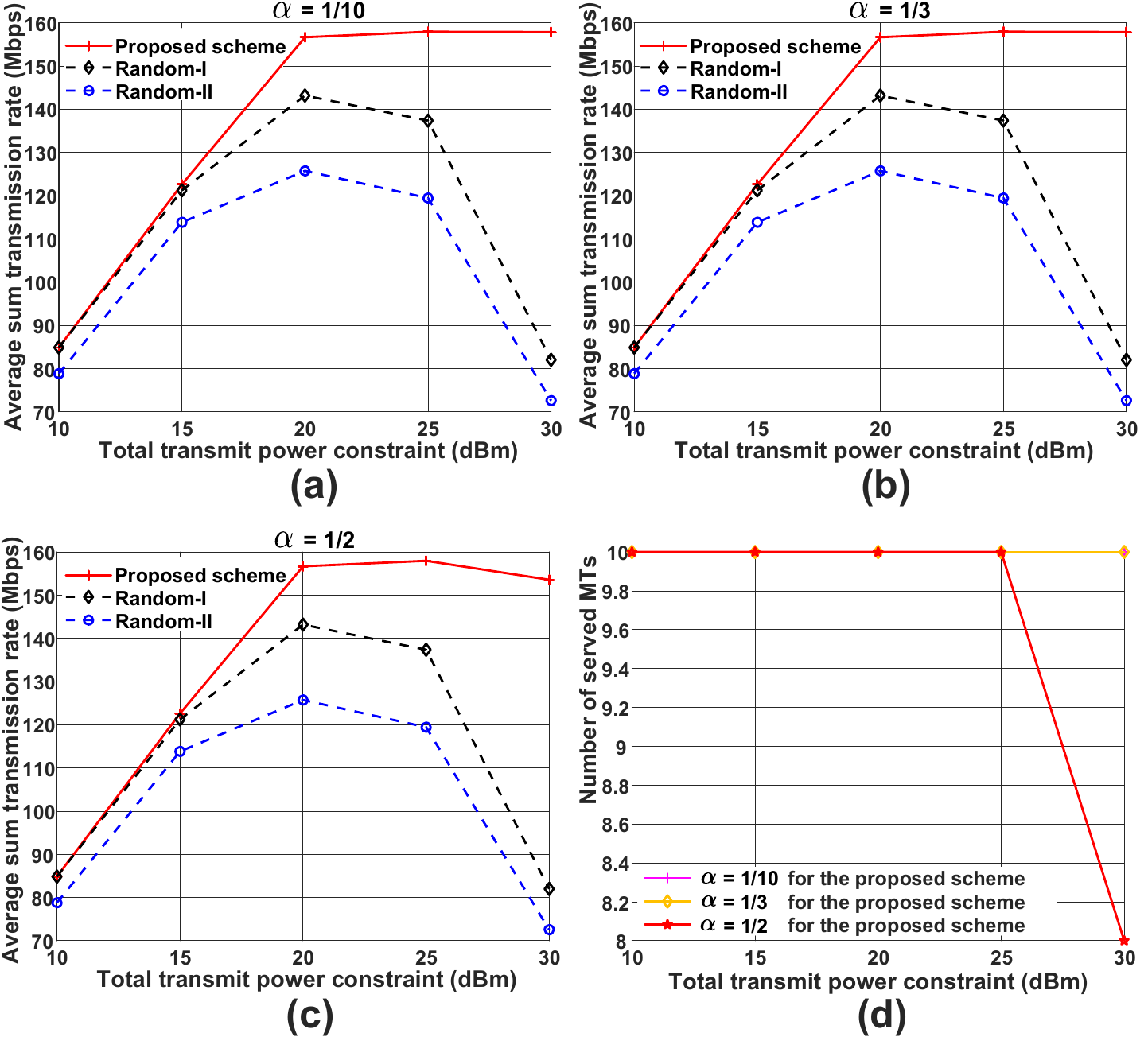}
		\caption{Performance comparison for $\alpha=1/10$, $1/3$, and $1/2$.}
		\label{fig5}
	\end{center}
\end{figure}

{Fig.~\ref{fig4} presents the average sum transmission rate of the terrestrial system
	achieved by spectrum sharing in the HSTN, 
	i.e., $\sum_{i=1}^{K_t} \sum_{j=1}^{K_s}  z^*_{i,j} R^*_{ij} B$, for $\alpha = 1/10$.
	Note that results in Fig.~\ref{fig4} are obtained based on average among $10$ random topologies for the HSTN.
	For comparison, two random schemes, i.e., Random-I and Random-II, are considered.
	Random-I is obtained based on the power allocation subproblem in~(\ref{eq9}) and random channel allocation for the terrestrial MTs.
	Random-II is obtained based on equal power allocation and random channel allocation for the terrestrial MTs.
	It can be seen from Fig.~\ref{fig4} that the proposed scheme achieves a significant performance gain
	over two random schemes, especially at relatively large total transmit power constraints.
	Furthermore, it can also be observed that 
	while the average sum transmission rate achieved by the proposed scheme increases rapidly from low to medium transmit power constraints,
	it gradually approaches a stable value at large transmit power constraints.
	It can be explained by the fact that transmit power of the BSs for terrestrial MTs can't keep growing
	along with the total transmit power constraint due to the constraints on CCI to satellite MTs.
	As for the two random schemes, a rapid decrease of the achieved average sum transmission rate 
	can be seen from medium to large transmit power constraints.
	It is because the number of served terrestrial MTs decreases rapidly along with the improvement of service quality requirements
	indicated by $\underline{R}_{i} = \alpha R^*_{ij} B $, $i=1,...,K_t$.}

{Fig.~\ref{fig5} compares performance of the proposed scheme for $\alpha=1/10$, $1/3$, and $1/2$.
	In the figure, (a)-(c) show the achieved average sum transmission rate, and (d) shows the number of terrestrial MTs
	that are served with service quality assurance by the proposed scheme.
	It can be observed that an obvious difference of (c) from (a) and (b) is that 
	the achieved average sum transmission rate by the proposed scheme deceases from $P_i=25$dBm to $P_i=30$dBm.
	It can be explained by the fact that the number of served terrestrial MTs decreases from $10$ to $8$
	when the service quality requirement $\underline{R}_{i} = \alpha R^*_{ij} B $ improves 
	from $1/10R^*_{ij} B$ and$1/3R^*_{ij} B$ to $1/2R^*_{ij} B$, as indicated by (d).}

\section{Conclusions}
{In this paper, we have addressed the problem of double-target collaborative spectrum sharing 
	for HSTNs based on large-scale CSI. 
	By introducing user-centric channel pools, 
	an efficient spectrum sharing scheme is proposed based on joint power and channel allocation.
	With potentials to achieve a significant performance gain, 
	the proposed scheme is rather promising for practical applications in HSTNs.}


\end{document}